\documentclass[pra,floatfix,showpacs,amsmath,twocolumn]{revtex4}

\usepackage{amssymb}
\usepackage[mathscr]{eucal}
\usepackage[dvips]{graphicx}

\begin{document}

\newcommand{\mywidth}{0.485\textwidth}
\newcommand{\tr}{\textrm{tr}}
\newcommand{\spec}{\textrm{spec}}
\newcommand{\Span}{\textrm{span}}
\newcommand{\ket}[1]{| \, #1 \, \rangle}
\newcommand{\bra}[1]{\langle \, #1 \, |}
\newcommand{\proj}[1]{\ket{#1} \bra{#1}}
\newcommand{\Proj}[2]{\ket{#1}_{#2} \bra{#1}}
\newcommand{\scal}[2]{\bra{#1} \, #2 \, \rangle}
\newcommand{\expect}[1]{\langle #1 \rangle}
\newcommand{\Expect}[2]{\langle #1 \rangle_{#2}}
\newcommand{\HH}{\mathcal{H}}
\newcommand{\adj}[1]{#1^{\dagger}}
\newcommand{\binomial}[2]{{#1 \choose #2}}
\newcommand{\round}[1]{\left[ #1 \right]}
\newcommand{\roundceil}[1]{\left\lceil #1 \right\rceil}
\newcommand{\roundfloor}[1]{\left\lfloor #1 \right\rfloor}
\newcommand{\Varepsilon}{\mathscr{E}}
\newcommand{\EE}{\mathcal{E}}
\newcommand{\arccot}{\textrm{arccot}}
\newcommand{\sgn}{\textrm{sgn}}
\newcommand{\ketr}{| \rightarrow \rangle}
\newcommand{\ketl}{| \leftarrow \rangle}
\newcommand{\ketu}{| \uparrow \, \rangle}
\newcommand{\ketd}{| \downarrow \, \rangle}
\newcommand{\brar}{\langle \rightarrow |}
\newcommand{\sx}[1]{\sigma_x^{(#1)}}
\newcommand{\sy}[1]{\sigma_y^{(#1)}}
\newcommand{\sz}[1]{\sigma_z^{(#1)}}
\newcommand{\splus}[1]{\sigma_{#1}^{+}}
\newcommand{\sminus}[1]{\sigma_{#1}^{-}}
\newcommand{\fpi}[2]{\left( \frac{2 \pi #1}{#2} \right) }
\newcommand{\omicron}[1]{\textrm{O} \left[ \frac{1}{#1} \right] }
\newcommand{\order}[1]{\left( \frac{1}{#1} \right)}

\title{Adiabatic Time Evolution in Spin-Systems}

\author{V. Murg}
\author{J. I. Cirac}
\affiliation{Max-Planck-Institut f\"ur Quantenoptik,
Hans-Kopfermann-Str. 1, Garching, D-85748, Germany.}

\date{\today}

\begin{abstract}
Adiabatic processes in the quantum Ising model and the anisotropic
Heisenberg model are discussed. The adiabatic processes are
assumed to consist in the slow variation of the strength of the
magnetic field that environs the spin-systems. These processes are
of current interest in the treatment of cold atoms in optical
lattices and in Adiabatic Quantum Computation. We determine the
probability that, during an adiabatic passage starting from the
ground state, states with higher energy are excited.
\end{abstract}

\maketitle

\section{Introduction}

The analysis of ground state properties of spin systems in the
vicinity of quantum phase transitions is a contemporary issue in
statistical physics and condensed matter physics. These systems
display a rich variety of phenomena which explain many exotic
properties of materials at very low temperature. Very recently, it
has been recognized that cold atoms in optical lattices can be
very well described in terms of certain spin
Hamiltonians~\cite{jaksch98,greiner02}. These systems can be very
well controlled at the quantum level, which may allow us to
observe physical phenomena predicted for spin systems and that
have, so far, not been observed with other systems. Moreover, they
may allow to build a quantum simulator which may shed some light
in unresolved issues in condensed matter physics~\cite{jane02}.

In the strong interacting regime and at sufficiently low
temperature, atoms confined in optical lattices tend to distribute
in the so--called Mott phase; that is, under the appropriate
conditions each lattice site is occupied by a single atom. Each
atom still interacts with the nearest neighbor via virtual
tunneling, which gives rise to an effective interaction between
the internal levels of neighboring atoms. Under certain
conditions, these interactions can be described in terms of the
Ising Model or the anisotropic Heisenberg Model with a transverse
magnetic field~\cite{duan02,ripoll03,dorner02}. Changing the laser
parameters amounts to varying the parameters of these models. Both
models display quantum phase transitions~\cite{sachdev} for
certain values of these parameters. Thus, with these systems it is
possible to investigate the dynamics of quantum phase transitions
by, for example, adiabatically changing these laser properties
\footnote{In fact, one can use this transition to prepare and
manipulate Schr\"odinger cat states which may be used to store a
qubit.}. An important question in this case is to what extent the
process can remain adiabatic since near a phase transtion the
energy levels tend to get closer as the number of particles
increases~\cite{dorner02}. In this paper we will study this
problem for both Hamiltonians. For the exactly solvable Ising
Model, the adiabatic process will be investigated in detail and
analytical results for the excitation probability will be given.
For the yet unsolved anisotropic Heisenberg Model, the adiabatic
process will be investigated by means of perturbation theory.

Another subject in which the results about adiabatic processes in
spin-systems are relevant is Adiabatic Quantum
Computation~\cite{farhi00,childs01,dam02}. In Adiabatic Quantum
Computation, solutions to mathematical and physical problems are
obtained by simulating adiabatic processes on quantum
computers~\cite{feynman82,lloyd96,abrams97,wu02}. An interesting
problem related to Adiabatic Quantum Computation is the
investigation of the ground state of spin-systems. This problem is
known to be difficult to be solved on classical computers, since
the required resources in time and space scale
\emph{exponentially} with the number of particles the system
consists of. Because of this exponential scaling, the ground state
properties can only be obtained for spin-systems consisting of a
few dozens of particles. In addition to that, analytical
calculations are rarely possible, such that ground state
properties of many spin-systems are still unknown. The question
whether the ground state of spin-systems can be investigated
efficiently by means of Adiabatic Quantum Computation is related
to the question how slowly parameters of the system must be
changed such that processes remain adiabatic. This question will
be answered in detail for the quantum Ising model. Even though
this model is exactly solvable and its ground state is known, the
results about this model shed light on the efficiency of adiabatic
quantum algorithms investigating the ground state of more
complicated spin-systems. In the case of the anisotropic
Heisenberg model, statements about the efficiency will be made in
regimes amenable to perturbative treatment.

The investigation of the ground state of spin-systems by Adiabatic
Quantum Computation is an example for a quantum algorithm that is
feasible with current technology: This algorithm only requires a
low number of qubits, of the order of~$50$, in order to exceed
classical computations. In addition, as we will show, the
algorithm is sufficiently robust against errors and imperfections,
such that no quantum error correction codes are necessary. On top
of that, the algorithm can be implemented by any experimental
realization of a Universal Quantum Simulator~\cite{lloyd96}, like
that based on optical lattices and arrays of
microtraps~\cite{jane02}.

The paper is outlined as follows: In the first part, general
statements about the quantum Ising model and the anisotropic
Heisenberg model are made. In the second part, the investigation
of the ground state of spin-systems by Adiabatic Quantum
Computation is discussed. In the third and forth part, adiabatic
processes in the quantum Ising model and the anisotropic
Heisenberg model are studied. The effects of perturbations on
these adiabatic processes are discussed in the last part.

\section{Preliminaries}

\subsection{Ising Model} \label{sec:ising}

The quantum Ising chain~\cite{pfeuty70,sachdev} is a perfectly
suited model for investigating adiabatic processes because it is
exactly solvable and shows a quantum phase transition. The
Hamiltonian that describes this model reads
\begin{equation} \label{eqn:hising}
    H = - J \sum_{i=0}^{N-1} \left( \sz{i} \sz{i+1} + g \, \sx{i} \right).
\end{equation}
The chain is assumed to be cyclic, i.e. $\sigma_{\xi}^{(j+N)}
\equiv \sigma_{\xi}^{(j)}$. The Pauli operators $\sx{i}$, $\sy{i}$
and $\sz{i}$ describe the spin of the $i^{th}$ particle in the
chain. The dimensionless variable~$g$ determines the strength of
the transverse magnetic field. The parameter~$J$ is considered as
a positive quantity that fixes the microscopic energy scale.
$N$~denotes the number of particles in the chain.

The diagonalization of Hamiltonian~(\ref{eqn:hising}) can be
performed by means of Jordan-Wigner transformation~\cite{lieb61}.
This transformation maps the Pauli operators~$\sigma_{x}^{(i)}$,
$\sigma_{y}^{(i)}$ and $\sigma_{z}^{(i)}$ on Fermi operators. The
details of the diagonalization procedure can be gathered from
appendix~\ref{sec:diagonalization}. After diagonalization, the
Hamiltonian of the quantum Ising model assumes a simple form,
namely,
\begin{displaymath}
    H/J = g N + \left\{
    \begin{array}{lcl}
        \sum_{k=0}^{N-1} \Lambda_k \left( \adj{\eta_k} \eta_k - \frac{1}{2}
        \right), &  & \hat{N} \textrm{ even}\\
        \sum_{k=0}^{N-1} \bar{\Lambda}_k \left( \adj{\bar{\eta}_k} \bar{\eta}_k - \frac{1}{2}
        \right), &  & \hat{N} \textrm{ odd}\\
    \end{array}
    \right\}
    ,
\end{displaymath}
with $\hat{N}$ denoting the number of fermions. The variables
$\Lambda_k$ and $\bar{\Lambda}_k$ denote energies of single
fermions and the $\eta$'s are fermionic operators.

\begin{figure}[htb]
    \begin{center}
        \includegraphics[width=\mywidth]{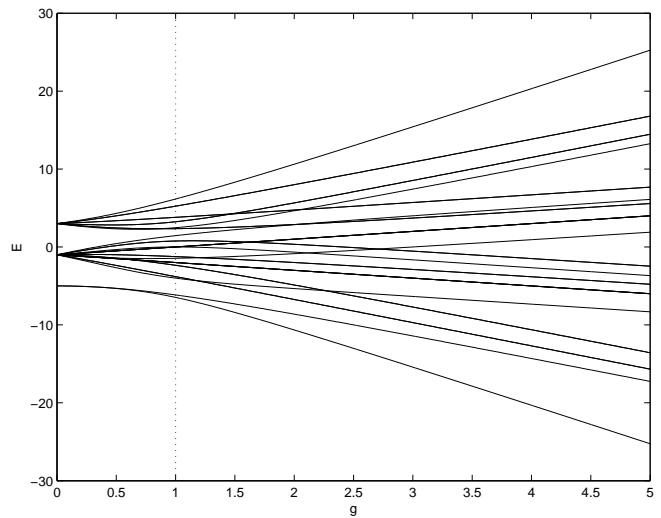}
    \end{center}
    \caption{Spectrum of an Ising chain consisting of $5$ particles ($N=5$)
        as a function of the transverse magnetic field $g$. The dotted line
        marks the critical point at which the quantum phase
        transition takes place.}
    \label{fig:ising:spec}
\end{figure}

The eigenvalues of Hamiltonian~(\ref{eqn:hising}) as functions of
the transverse field~$g$ are visualized in
figure~\ref{fig:ising:spec}. It is easily shown that the ground
state energy as a function of~$g$ is non-analytic at the point
$g=1$ in the thermodynamic limit. This non-analytic point is an
indication for a quantum phase transition~\cite{sachdev} at the
\emph{quantum critical point} $g=1$. This quantum phase transition
separates two phases at zero temperature: an ordered phase
for~$g<1$ and a disordered phase for~$g>1$. In the ordered phase,
the interaction between the particles aligns the spins parallel or
antiparallel to the crystalline axes. In the disordered phase, the
external magnetic field induces spin-flips that destroy this
order.

The symmetries that Hamiltonian~(\ref{eqn:hising}) possesses are
the $Z_2$-symmetry and the translational symmetry. The
$Z_2$-symmetry corresponds to a reflection of the spin-vectors at
the $x$-axes. It is generated by the operator
\begin{displaymath}
    Z_2 = \prod_{i=0}^{N-1} \sx{i}.
\end{displaymath}
This operator possesses two different eigenvalues, $1$~and~$-1$,
such that eigenstates can be classified by two different
$Z_2$-symmetries. In terms of Fermi operators, states with an even
number of fermions possess $Z_2$-symmetry~$1$ and states with an
odd number of fermions possess $Z_2$-symmetry~$-1$. The
translational symmetry is generated by the translation-operator
which has the property to right-shift all product states:
\begin{displaymath}
    T \ket{z_1} \cdots \ket{z_N} =
    \ket{z_N} \ket{z_0} \cdots \ket{z_{N-1}}
\end{displaymath}
This operator possesses $N$ different eigenvalues~$e^{-i 2 \pi
n/N}$ ($n=0,\ldots,N-1$), such that the eigenstates can be
classified by $N$ different translational symmetries.

\subsection{Heisenberg Model} \label{sec:heisenberg}

The anisotropic Heisenberg
chain~\cite{sutherland70,baxter71,kurmann81} is characterized by
anisotropic internal interactions. The Hamiltonian that defines
this model reads
\begin{eqnarray} \label{eqn:hheisenberg}
    H = - J \sum_{i=0}^{N-1}
    \big\{ \Delta_x \sx{i} \sx{i+1} & + & \Delta_y \sy{i} \sy{i+1} +\\
    + \Delta_z \sz{i} \sz{i+1} & + & g \, \sx{i} \big\} \nonumber.
\end{eqnarray}
The chain is assumed to be cyclic. The properties of the
interactions between the particles of the chain are described by
the anisotropy-parameters $\Delta_x$, $\Delta_y$ and $\Delta_z$
and the strength of the external magnetic field is determined by
the parameter~$g$.

The anisotropic Heisenberg model possesses the same symmetries as
the quantum Ising model: the $Z_2$-symmetry and the translational
symmetry. Thus, eigenstates can be classified by two different
$Z_2$-symmetries ($1$~and~$-1$) and by~$N$ different translational
symmetries ($e^{-i 2 \pi n/N}$ with $n=0,\ldots,N-1$).

\begin{figure}[tb]
    \begin{center}
        \includegraphics[width=\mywidth]{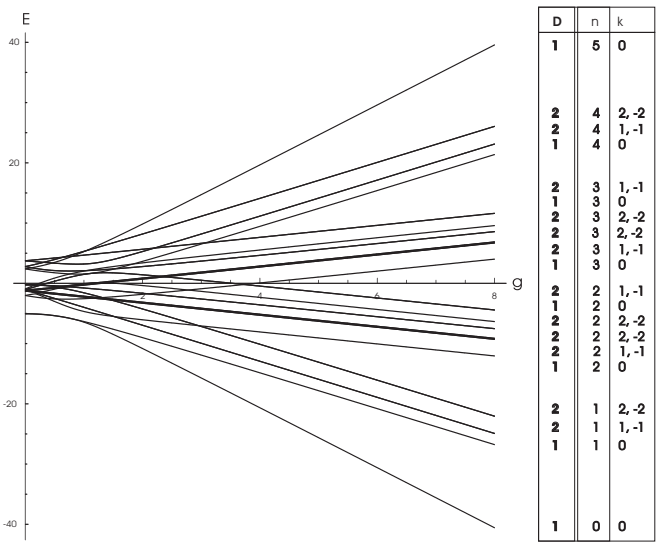}
    \end{center}
    \caption{Spectrum of the anisotropic Heisenberg model as a function of the external magnetic field $g$.
        Assumptions: $N=5$, $\Delta_x=0.1$, $\Delta_y=0.3$ and $\Delta_z=1$;
        The table on the right hand side gives informations about the degeneracy $D$, the group-index $n$ and
        the translational-symmetry $k$ of each level.}
    \label{fig:heisenberg:spec}
\end{figure}

Another possibility to classify the eigenstates is to classify
them according to their behavior in a very strong magnetic field. In
figure~\ref{fig:heisenberg:spec} the energies of the eigenstates
are plotted as functions of the field-strength~$g$. From this
figure it can be gathered that the eigenstates cluster in~$N+1$
groups because of the Zeeman-shifts. All states in each group have
the property that they tend to the same state in the limit $g \to
\infty$. In the $n^{th}$ group, for example, all states tend to a
state that has all but $n$~spins aligned in $x$-direction and
$n$~spins pointing in $-x$-direction. All eigenstates in the
$n^{th}$ group can be classified by a group-index~$n$. This
group-index is related to the $Z_2$-symmetry of the states. It can
be shown that states with an even group-index possess the
$Z_2$-symmetry~$1$ and states with an odd group-index possess the
$Z_2$-symmetry~$-1$.

In order to take into account that several eigenstates with
group-index $n$ and translational symmetry $k$ may exist, a third
index $d$ is required to distinguish eigenstates with equal values
for $n$ and $k$. In summary, all eigenstates of $H$ can be
identified by the three indices $n$, $k$ and $d$. They will be
denoted as
\begin{displaymath}
    \ket{\psi_{n k d}},
\end{displaymath}
where $n \in \{ 0,\ldots,N \}$ and $k \in \{ 0,\ldots,N-1 \}$.
Evidently, the kets $\ket{\psi_{n k d}}$ are common eigenstates
of~$H$, $Z_2$ and~$T$.

\section{Quantum simulations of spin systems}

Before studying adiabatic processes in spin-systems, a relevant
application shall be discussed: the investigation of the ground
state of spin-systems with a quantum computer. This investigation
could be performed by means of Adiabatic Quantum
Computation~\cite{farhi00,childs01,dam02}.

The scheme of Adiabatic Quantum Computation reads as follows:
First, the quantum computer is prepared in the ground state of a
simple \emph{beginning Hamiltonian}, the ground state of which is
known. Then, using the Universal Quantum Simulator~\cite{lloyd96},
a time evolution according to a time-dependent Hamiltonian~$H(t)$
is simulated which adiabatically interpolates between the simple
\emph{beginning Hamiltonian} and a complicated \emph{problem
Hamiltonian}, the ground state of which shall be investigated.
Because of the adiabatic variation, the quantum computer always
stays in the ground state of the time-dependent Hamiltonian~$H(t)$
and is finally prepared in the ground state of the complicated
\emph{problem Hamiltonian}. Finally, by appropriate measurements,
information about the ground state of the \emph{problem
Hamiltonian} is obtained.

Evidently, in the case of spin-systems, the \emph{problem
Hamiltonian} is the Hamiltonian that describes the spin-system of
interest. The ground state properties of this Hamiltonian are
usually interesting as functions of the strength~$g$ of an
external magnetic field. Thus, it is obvious to lay out the
algorithm such that the field-strength~$g$ is adiabatically varied
and tunes between the simple \emph{beginning Hamiltonian} and the
\emph{problem Hamiltonian}. The \emph{beginning Hamiltonian} can
either be the Hamiltonian of the spin-system for~$g=0$ or the
Hamiltonian of the spin-system for $g \to \infty$. The algorithm
then consists either in preparing the quantum computer in the
ground state for~$g=0$ and adiabatically increasing the
field-strength~$g$ or in preparing the quantum computer in the
ground state for $g \to \infty$ and adiabatically decreasing the
field-strength. The ground state for~$g \to \infty$ is simple
since it has all spins aligned in the direction of the external
field. Whether the ground state for~$g=0$ can be prepared or not
depends on the spin-system under study. In fact, many models exist
that are solvable for~$g=0$ and, in the case of these models,
$g=0$ can be chosen as a starting point.

The duration~$T$ of the algorithm is related to the change rate of
the parameter~$g$. The parameter~$g$ must be changed slowly
enough, such that the time evolution remains adiabatic. The change
rate can be determined mathematically by means of the Adiabatic
Theorem~\cite{born28,kato50,friedrichs55}. This theorem deals with
the solution of the Schr\"odinger equation in the case of a
time-dependent and slowly varying Hamiltonian. The statements of
the Adiabatic Theorem are the following: In the limit $T \to
\infty$, the system always stays in the ground state. Thus,
eigenstates of the \emph{beginning Hamiltonian} are mapped with
certainty on eigenstates of the \emph{problem Hamiltonian} and
there is only a change in the phase. In reality, however, the
duration~$T$ of the time evolution is finite. Thus, the question
that has to be answered is how large the duration~$T$ must be,
such that, with a high probability, the system stays in the ground
state. In other words, the probability that eigenstates with
higher energy are excited must be negligible.

A rough criterion on the duration~$T$ that guarantees that the
excitation probability is negligible reads~\cite{messiah}
\begin{equation} \label{eqn:adiabatic:t}
    T \gg \frac{\Varepsilon}{\Delta^2}.
\end{equation}
$\Varepsilon$ is thereby a quantity that depends on the derivative
of the Hamiltonian~$H(s)$ with respect to~$s$ and $\Delta$ denotes
the minimum energy difference between the ground state and the
first excited state of~$H(s)$. Usually, $\Varepsilon$ scales
polynomially with the number of particles, such that the
efficiency mainly depends on the behavior of the minimum energy
difference $\Delta$ as a function of the particle number~$N$. The
energy difference usually reaches its minimum at an avoided
level-crossing. At an avoided level-crossing, the ground state and
the first excited state approach as the number of particles
increases, such that the duration of the algorithm will always
increase as the number of particles increases. The efficiency of
the algorithm is then determined by the velocity with that the
first two states approach. If, on the one hand, the first two
states approach exponentially fast with a growing particle number,
the algorithm is not efficient. If, on the other hand, the two
states approach polynomially fast, the algorithm can be considered
to be efficient.


What is seen from this discussion is that knowledge of the
spectrum of the Hamiltonian~$H(s)$ is required in order to make
statements about the duration of the algorithm. However, the
spectrum is not known, in general. If the spectrum was known, the
ground state of the \emph{problem Hamiltonian} would be known, as
well, and no quantum computer would be required to calculate it.
Thus, the only thing that can be done is to test the quantum
algorithm on spin-systems that are solvable and see whether it is
efficient or not, or to try to find estimations of the spectrum
and use them to approximate the duration of the algorithm.

Because of these difficulties in the mathematical determination of
the duration of the algorithm, a simple experimental method is
desirable that can be used to check whether a chosen change-rate
is sufficient or not. An experimental method could look like this:
First, the quantum computer is prepared in the ground state of the
\emph{beginning Hamiltonian}, e.g. the Hamiltonian of the
spin-system for $g=0$. Then, the field-strength~$g$ is increased
with a chosen change rate up to a desired value and decreased
again with the same change rate. At the end, a measurement in the
Eigenbasis of the \emph{beginning Hamiltonian} is performed (which
is known). From this measurement it can be deduced whether the
quantum computer is still in the ground state or whether levels
with higher energy have been excited. If the quantum computer is
still in the ground state, the time evolution was adiabatic and
the change rate was low enough. Otherwise, the change rate must be
decreased and the experimental check must be performed once again.

\section{Adiabatic process in the Ising Model} \label{sec:adising}

The adiabatic process under study consists in the slow variation
of the strength of the magnetic field that environs the Ising
chain, i.e. the slow variation of parameter~$g$ in
Hamiltonian~(\ref{eqn:hising}). This variation leads to
excitations of states that possess the same symmetry as the
initial state of the system (which is assumed to be the ground
state).

\begin{figure}[t]
    \begin{center}
        \includegraphics[width=\mywidth]{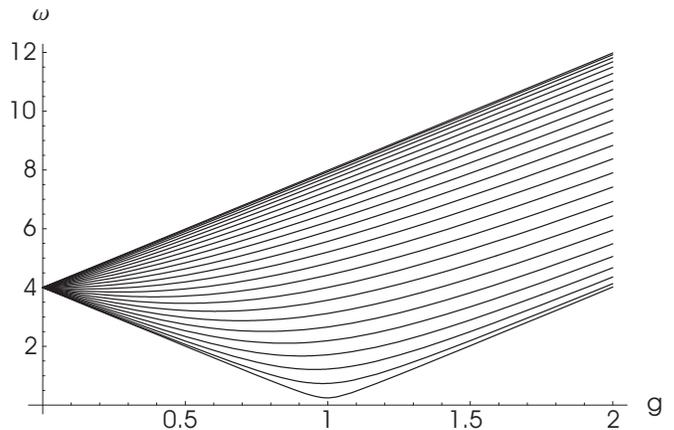}
    \end{center}
    \caption{Plot of the excitation energies of the states $\ket{n}$ as functions of $g$.
        The assumed number of particles $N$ is $51$.}
    \label{fig:myalgising:wn0}
\end{figure}

The symmetries of the problem are, as discussed in
section~\ref{sec:ising}, the $Z_2$-symmetry and the translational
symmetry. $Z_2$-symmetry imposes that only states consisting of an
even number of fermions are excited. Translational symmetry makes
further restrictions. The states that are excited in first order
can be identified by means of the Adiabatic
Approximation~\cite{messiah}. They read
\begin{displaymath}
    \ket{n} \equiv \adj{\eta_n} \adj{\eta_{-n}}
    \ket{vac},
\end{displaymath}
where $n = 1, \ldots, \lfloor \frac{N-1}{2} \rfloor$. The energies
of these states with respect to the ground state energy are
plotted in figure~\ref{fig:myalgising:wn0} as functions of the
field-strength~$g$. This figure shows that the energy difference
between the lowest excited state and the ground state tends to a
minimum at the critical point~$g=1$. The minimum energy difference
thereby amounts to $\frac{4 \pi}{N}$ for $N \gg 1$. This
observation already allows to make a qualitative statement about
the duration of an adiabatic time evolution in the vicinity of the
critical point: Since, according to
formula~(\ref{eqn:adiabatic:t}), the duration scales with
$\frac{1}{\Delta^2}$, where $\Delta$ is the minimal energy
difference between the lowest exited state and the ground state,
the duration can be expected to scale with $N^2$, i.e. with the
square of the number of particles. This has the consequence that,
for example, the investigation of the ground state of the quantum
Ising model in the vicinity of the critical point by means of
Adiabatic Quantum Computation will be efficient.

These qualitative results can be clarified by calculating the
excitation probabilities~$p_{0 \to n}$ for all states~$\ket{n}$.
This can be done analytically by making use of the Adiabatic
Approximation~\cite{messiah}. The details of the analytical
calculation can be gathered from
appendix~\ref{sec:calculation_ising}. Another way to calculate the
excitation probabilities~$p_{0 \to n}$ consists in numerically
solving the Heisenberg equations for the Fermi-operators~$\eta_k$.
Using this solution, it is simple to calculate expectation values
of arbitrary observables after the adiabatic process. If it is
assumed that the final state of the system is a superposition of
the ground state and the $2$-fermion states~$\ket{n}$ only, the
excitation probabilities~$p_{0 \to n}$ can be calculated as well,
since they are expressible in terms of expectation values of the
operators $\adj{\eta_n} \adj{\eta_{-n}} \eta_{-n} \eta_n$.

\begin{figure}[t]
    \begin{center}
        \includegraphics[width=0.45\textwidth]{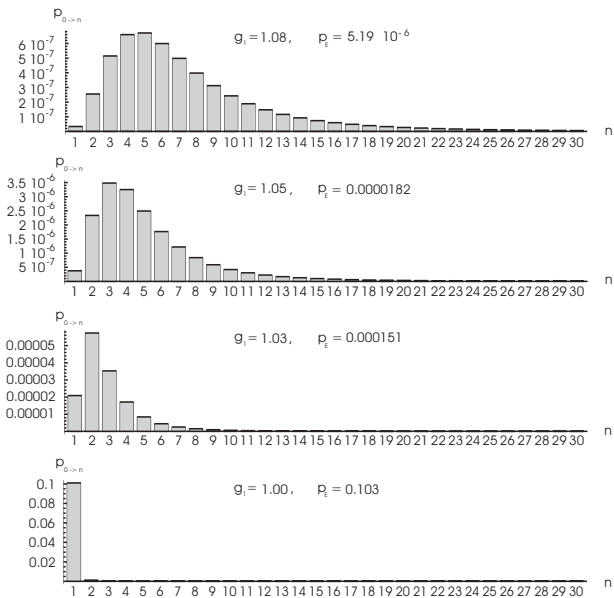}
    \end{center}
    \caption{
        Excitation probability~$p_{0 \to n}$ of the levels $\ket{n}$ in the case of $501$ particles ($N=501$).
        The magnetic field is assumed to be varied from $g_0=5$ to $g_1$ with
        change rate $\dot{g}$ equal to~$-0.0001$. The
        indices~$n$ are reshuffled such that states with lower
        index have lower energy.
        }
    \label{fig:myalgising:p0n500}
\end{figure}

In figure~\ref{fig:myalgising:p0n500}, the excitation
probabilities~$p_{0 \to n}$ are plotted for an Ising chain
consisting of~$501$ particles. The adiabatic process is assumed to
consist in the linear decrease of the magnetic field-strength~$g$
with change rate $\dot{g}=-0.0001$ from~$g_0=5$ to~$g_1=1.08$,
$1.05$, $1.03$ and $1.0$ respectively. It can be observed that, if
$g_1$ is in the vicinity of the critical point, the lowest state
is excited with maximal probability.

\begin{figure}[t]
    \begin{center}
        \includegraphics[width=\mywidth]{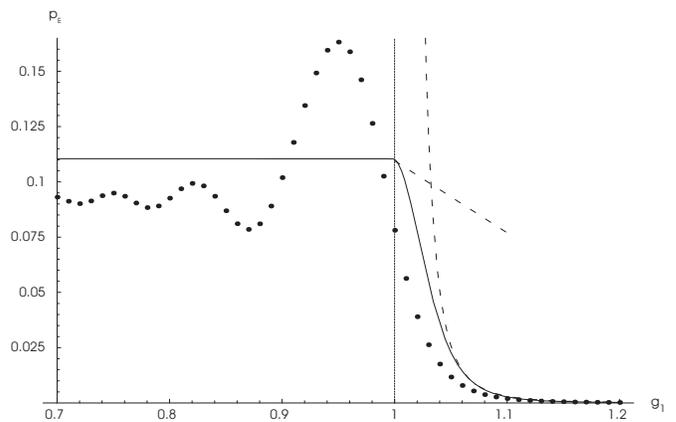}
    \end{center}
    \caption{
        Excitation probability $p_E$ as a function of the final
        field strength~$g_1$.
        Assumptions: $51$ particles ($N=51$); linear variation of $g$ from $g_0=5$ to
        $g_1$ with change rate $\dot{g} = -0.01$.
        \emph{Solid line}: Sum of all terms $p_{0 \to n}$;
        \emph{Dotted line}: Numerical results (see text);
        \emph{Dashed lines}: Analytical estimations (formulas (\ref{eqn:myalgising:peapproxa}) and
        (\ref{eqn:myalgising:peapproxb})).
        }
    \label{fig:myalgising:pen50g1}
\end{figure}

The overall excitation probability~$p_E$ is obtained by summing
the excitation probabilities~$p_{0 \to n}$ of the respective
states. Analytical estimations for the overall excitation
probability are obtained in three different regimes:
\newline (1) $g_0 \gg 1$, $g_1>1$ and $N (g_1-1) \gg 1$
\newline (2) $g_0 \gg 1$, $g_1>1$ and $N (g_1-1) \ll 1$
\newline (3) $g_0 \gg 1$ and $g_1 \leq 1$

\begin{itemize}
\item In regime~(1), the analytical expression for~$p_E$ reads
\begin{equation} \label{eqn:myalgising:peapproxa}
    p_E \lesssim
    \frac{\hbar^2}{2^8 J^2}
    \frac{(g_0-g_1)^2}{T^2}
    \frac{1}{(g_1^2-1)^3} N,
\end{equation}
such that the duration~$T$ of the adiabatic process will scale
with the square root of the number of particles~$N$.

\item In regime~(2), the overall excitation probability~$p_E$ can
be written as
\begin{equation} \label{eqn:myalgising:peapproxb}
     p_E \lesssim \frac{\hbar^2}{2^6 \pi^4 J^2}
    \frac{(g_0-g_1)^2}{T^2}
    (4-3g_1) N^4.
\end{equation}
Thus, the duration~$T$ of the adiabatic process will show the
previously mentioned $N^2$-scaling.

\item In regime~(3), the overall excitation probability~$p_E$ is
constant as a function of~$g_1$ (for a fixed change-rate $\dot{g}
\equiv (g_0-g_1)/T$), namely,
\begin{equation} \label{eqn:myalgising:peapproxc}
     p_E \lesssim \frac{\hbar^2}{2^6 \pi^4 J^2}
    \frac{(g_0-g_1)^2}{T^2}
    N^4
\end{equation}
and the duration~$T$ of the adiabatic process will show a
quadratic scaling with the number of particles~$N$, just like in
regime~(2).
\end{itemize}

\begin{figure}[htb]
    \begin{center}
        \includegraphics[width=0.48\textwidth]{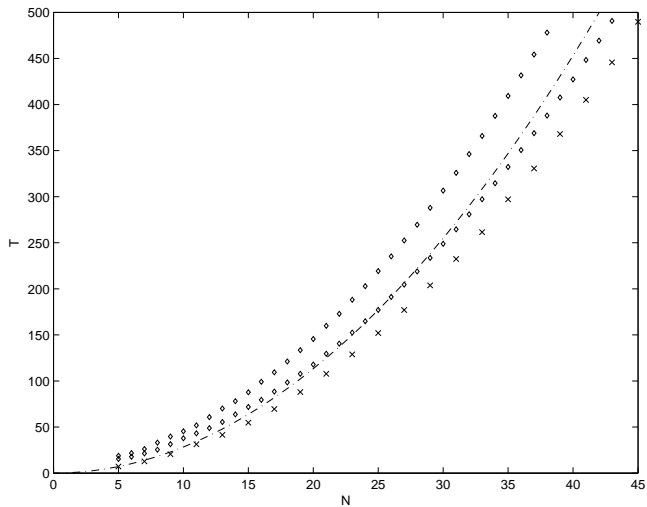}
    \end{center}
    \caption{Duration~$T$ of the adiabatic process as a function of the particle-number~$N$.
        Assumptions: $g$ is linearly decreased from~$g_0=5$ to~$g_1=0$;
        the change-range~$\dot{g}$ is adjusted such that the
        excitation probability~$p_E$ always amounts to~$0.05$;
        \emph{Crosses}: numerical results (see text);
        \emph{Dash-dotted line}: analytical result~(\ref{eqn:myalgising:peapproxc}), i.e. $T
        \approx 0.2832 N^2$.
        \emph{Diamonds}: results presented in~\cite{dorner02} for the free-end chain.
        }
    \label{fig:myalgising:TN}
\end{figure}

In figure~\ref{fig:myalgising:pen50g1}, the overall excitation
probability~$p_E$ after the decrease of~$g$ from~$g_0=2$ down
to~$g_1$ is shown as a function of the final field-strength~$g_1$.
The change rate~$\dot{g}$ is assumed to be fixed at~$-0.01$ and
the number of particles is taken to be~$51$. The solid line
represents the sum of the terms $p_{0 \to n}$ that were calculated
by means the Adiabatic Approximation. The dashed lines represent
the analytical estimations~(\ref{eqn:myalgising:peapproxa})
and~(\ref{eqn:myalgising:peapproxb}). The dots represent numerical
results obtained by solving the Heisenberg equations for the Fermi
operators. What can be observed is that the excitation probability
increases fundamentally as the critical point~$g=1$ is crossed.
This behavior is due to the fundamental approach of the ground
state and the lowest excited state at the critical point.

The duration~$T$ as a function of the number of particles~$N$ of
an adiabatic process consisting in the linear decrease of the
field-strength~$g$ from $g_0=5$ to $g_1=0$ can be gathered from
figure~\ref{fig:myalgising:TN}. The change-rate~$\dot{g}$ is
thereby adjusted such that the overall excitation
probability~$p_E$ is always equal to~$0.05$. The analytic
result~(\ref{eqn:myalgising:peapproxc}) is represented by the
dash-dotted line and the results obtained from numerically solving
the Heisenberg equations for the~$\eta_k$'s are represented by
crosses. In both cases, a quadratic scaling of the duration~$T$
with the number of particles~$N$ can be observed.

\begin{figure}[htb]
    \begin{center}
        \includegraphics[width=0.48\textwidth]{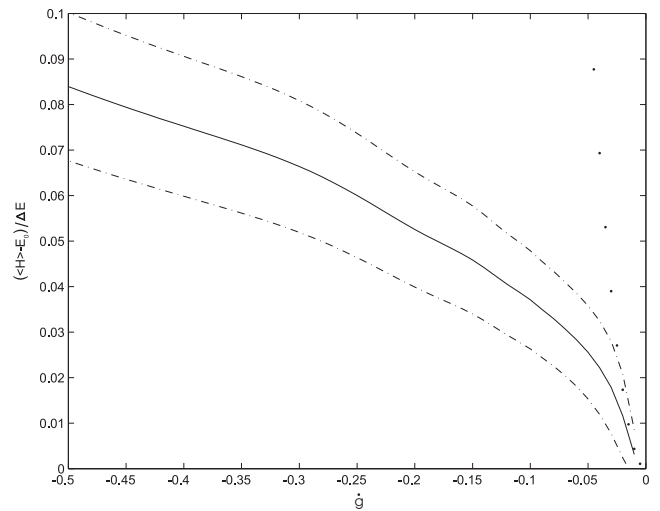}
    \end{center}
    \caption{Expectation value of $H$ relative to the width of the spectrum, i.e.
    $(\expect{H}-E_0)/\Delta E$, as a function of the
    change-rate~$\dot{g}$. Assumptions: $N=51$, $g_0=5$ and $g_1=0$;
    \emph{solid line}: numerical results (see text);
    \emph{dash-dotted line}: indicates variance of~$H$;
    \emph{dots}: results obtained using the Adiabatic Approximation.
    }
    \label{fig:myalgising:evh}
\end{figure}

Up to now it has been assumed that the field-strength is changed
sufficiently slowly, such that the Adiabatic
Approximation~\cite{messiah} is valid. Outside the adiabatic
regime, states with a fermion-number $4,6,8,\ldots$ are excited,
additionally to the $2$-fermion states~$\ket{n}$. These states
contribute to the overall excitation probability and must be taken
into account. However, even outside the adiabatic regime, the mean
energy of the system still remains very low. This can be gathered
from figure~\ref{fig:myalgising:evh}. In this figure, the mean
energy after the decrease of the field-strength~$g$ from $g_0=5$
down to $g_1=0$ is plotted as a function of the
change-rate~$\dot{g}$ for a~$51$-particle chain. The results about
the mean energy were obtained by numerically solving the
Heisenberg equations for the~$\eta_k$'s. As a consequence,
processes marginally outside the adiabatic regime prepare the
system in a state with very low temperature the properties of
which are still interesting to be investigated with a quantum
computer.

\section{Adiabatic process in the Heisenberg Model}

The adiabatic process in the anisotropic Heisenberg model is
treated similarly as in the previous section: Symmetries are used
to determine the states that are excited during the adiabatic
process. Since the adiabatic process starts from the ground state,
the excited states can be identified as the states that possess
the same symmetries as the ground state. Thus, using the notation
of section~\ref{sec:heisenberg}, the excited states are the states
with an even group-index~$n$ and translational symmetry~$k=0$. In
practice, only states from the $2^{nd}$~group are excited in first
order, namely,
\begin{displaymath}
    \ket{\psi_{2 0 d}}
\end{displaymath}
with $d$ ranging from $1$ to $\frac{N-1}{2}$ ($N$ is assumed to be
odd).

Using the previous considerations, it is straightforward to study
the efficiency of the algorithm in the case of the anisotropic
Heisenberg model. What is required to be done is to find
perturbative expansions of the ground state and the possible
excited states and to use the Adiabatic
Approximation~\cite{messiah} to calculate the excitation
probabilities. Of course, in order to be allowed to truncate the
perturbative expansions after a few lower-order terms, it must be
demand that the Hamiltonian consists of parts that are of a
different order of magnitude. Because of this, it is assumed that
the strength~$g$ of the external magnetic field is always much
larger than the strength of the internal interactions. The
adiabatic process therefore consists in adiabatically decreasing
the field strength~$g$ from~$g_0 \to \infty$ down to a certain
value~$g_1$, which is still much larger than the strength of the
internal interactions ($g_1 \gg \Delta_{\xi}$).

\begin{figure}[t]
        \begin{center}
            \includegraphics[width=0.38\textwidth]{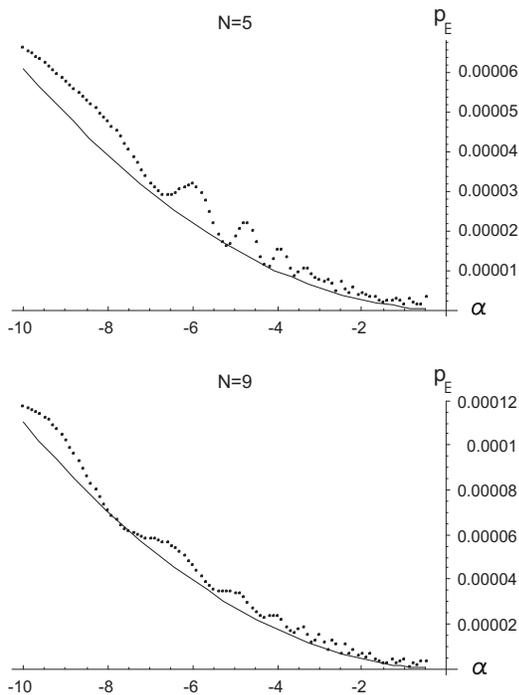}
        \end{center}
        \caption{
            Excitation probability $p_E$ as a function of the change rate
            $\alpha=(g_1-g_0)/T$ in the case of $5$- and $9$-particle chains.
            Assumptions: $\Delta_x=0.1$, $\Delta_y=0.3$, $\Delta_z=1$; variation of $g$ from $g_0=10$ to
            $g_1=5$;
            \emph{Dotted line}: Excitation probability obtained from a numerical simulation.
            \emph{Solid line}: Approximation~(\ref{eqn:myalgheisenberg:peapprox}) of the excitation probability in the limit $N \gg 1$.
        }
    \label{fig:myalgh:pea}
\end{figure}

The results that are obtained in this way are very similar to
results obtained for the quantum Ising model. Details about the
calculations can be gathered from
appendix~\ref{sec:calculation_heisenberg}. The analytical
expression that is obtained for the overall excitation
probability~$p_E$ reads
\begin{equation} \label{eqn:myalgheisenberg:peapprox}
    p_E \lesssim
    \frac{\hbar^2}{2^8 J^2} \left( \Delta_z - \Delta_y \right)^2
    \frac{(g_0-g_1)^2}{T^2}
    \frac{1}{g_1^6} N,
\end{equation}
such that the duration~$T$ of the adiabatic process will scale
with the square root of the number of particles~$N$. As a
consequence, it will be possible to investigate the ground state
of the anisotropic Heisenberg model efficiently by means of
Adiabatic Quantum Computation in the regime where the external
magnetic field is much stronger than the internal interactions.

The analytical expression~(\ref{eqn:myalgheisenberg:peapprox}) can
be compared to the exact result~(\ref{eqn:myalgising:peapproxa})
presented in section \ref{sec:adising} for the quantum Ising
model. As it is easily verified, the exact result is equal to
expression~(\ref{eqn:myalgheisenberg:peapprox}) in the limit $g_1
\gg 1$ if the parameters $\Delta_x$, $\Delta_y$ and $\Delta_z$ are
specified to $\Delta_x = 0$, $\Delta_y = 0$ and $\Delta_z = 1$.

\begin{figure}[t]
    \begin{center}
        \includegraphics[width=0.38\textwidth]{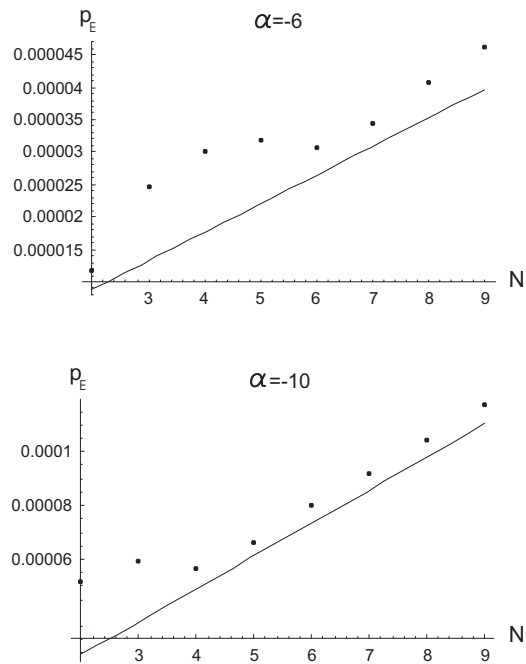}
    \end{center}
    \caption{
        Excitation probability $p_E$ as a function of the number
        of particles $N$ in the case of change rates $\alpha=-6$ and $\alpha=-10$.
        Assumptions: $\Delta_x=0.1$, $\Delta_y=0.3$, $\Delta_z=1$; variation of $g$ from $g_0=10$ to
        $g_1=5$;
        \emph{Dotted line}: Excitation probability obtain from a numerical simulation.
        \emph{Solid line}: Approximation~(\ref{eqn:myalgheisenberg:peapprox}) of the excitation probability in the limit $N \gg 1$.
    }
    \label{fig:myalgh:pen}
\end{figure}

Expression~(\ref{eqn:myalgheisenberg:peapprox}) can also be
compared to results obtained from numerical simulations. This is
shown in figures~\ref{fig:myalgh:pea} and~\ref{fig:myalgh:pen}. In
these figures, the excitation probability is plotted as a function
of the change rate~$\dot{g}$ and as a function of the
particle-number~$N$ respectively. The dots represent the results
from numerical simulations and the solid lines represent
evaluations of the analytical
formula~(\ref{eqn:myalgheisenberg:peapprox}). It is seen that the
numerical results for the excitation probability scale linearly
with the number of particles~$N$ and they scale quadratically with
the change rate~$\dot{g}$, as it is stated by
formula~(\ref{eqn:myalgheisenberg:peapprox}).

\section{Stability}

In this section, the implication of perturbations that accompany
the adiabatic process shall be discussed. These perturbations
occur, for example, if the adiabatic process is simulated on a
quantum computer and quantum gates are not implemented perfectly.
The problem that arises is that perturbations usually break the
symmetry of the original Hamiltonian, such that, in principle,
excitations of levels with very low energy may occur. The question
that has to be answered in this connection is whether these
excitations spoil previous results or not.

In the case of the quantum Ising model, the answer to this
question can be found using time-dependent perturbation theory
with respect to the perturbation. The perturbation is thereby
assumed to be given by
\begin{displaymath}
    V = J \varepsilon \, \sz{0}.
\end{displaymath}
If the adiabatic process consists in the slow decrease of the
field-strength~$g$ from $g_0 \gg 1$ to $g_1<1$, then, after
crossing the critical point, this perturbation leads to a strong
excitation of the (asymptotically degenerate) first excited state
and a very small excitation of states with higher energy.
Nevertheless, the mean energy of the system after excitation is
still very low compared to the width of the spectrum~$\Delta E$,
namely,
\begin{displaymath}
    \frac{\expect{H}-E_0}{\Delta E} = const \, \varepsilon^2.
\end{displaymath}
This term doesn't depend on the number of particles~$N$ and is
negligible for $\varepsilon \ll 1$ even for large~$N$. Thus, it
can be concluded that, even though an adiabatic process
accompanied by small perturbations will not leave the system in
the ground state, it will prepare the system in a state with very
low temperature. Such a low-temperature state also shows
properties which are interesting to be investigated with a quantum
computer.

In the case of the anisotropic Heisenberg model, the perturbation
term is assumed to be more general, namely,
\begin{displaymath}
    V = J \sum_{j=0}^{N-1} \varepsilon_j \, \vec{n}^{(j)} \cdot
    \vec{\sigma}^{(j)},
\end{displaymath}
with $\varepsilon_j \in \mathbb{R}$ and $|\vec{n}^{(j)}| = 1$. The
adiabatic process, however, is assumed to be restricted to the
regime where the external magnetic field is much stronger than the
internal interactions. The probability that the perturbation~$V$
causes an excitation during this process is then estimated~as
\begin{displaymath}
    p_V \lesssim \left( \frac{\varepsilon}{2 g_0} \right)^2
    N
\end{displaymath}
where~$\varepsilon$ is defined as
\begin{displaymath}
    \varepsilon := \sqrt{ \frac{1}{N} \sum_{j=0}^{N-1} \varepsilon_j^2
    \left| n_x^{(j)} + i n_y^{(j)} \right|^2 }.
\end{displaymath}
From this formula it can be read off that the
excitation-probability $p_V$ scales \emph{polynomially} with the
error-parameter $\varepsilon$ and the number of particles $N$.
Thus, small perturbations during the adiabatic process will not
have severe implications.

\section{Conclusions}

Summing up, adiabatic processes have been investigated in the
light of the simple, exactly solvable quantum Ising model and the
more complicated anisotropic Heisenberg model. The adiabatic
processes were assumed to consist in the slow variation of the
strength of the magnetic field that environs the spin-chains.

In the case of the quantum Ising model, the investigation could be
performed in detail and analytic results were obtained even for
processes that cross the quantum critical point. In the case of
the anisotropic Heisenberg model, adiabatic processes were studied
in regimes amenable to perturbative treatment. In both cases, the
duration of the adiabatic processes turned out to scale
polynomially with the number of particles the spin-systems consist
of.

The results that were obtained are relevant for the treatment of
bosons in optical lattices and for Adiabatic Quantum Computation.
In Adiabatic Quantum Computation, light is shed on the efficiency
of adiabatic quantum algorithms that investigate the ground state
of spin-systems. In the treatment of optical lattices, insight is
gained into the dynamics of quantum phase transitions.

As an outlook on future work, adiabatic processes in more
complicated, yet unsolved models, such as higher-dimensional
spin-systems or spin-glasses, could be studied, such that
information is gained about quantum phase transitions in these
models and about the efficiency of adiabatic quantum algorithms
investigating the ground state of these models.

\appendix

\section{Diagonalization of the Quantum Ising-Hamiltonian} \label{sec:diagonalization}

Jordan-Wigner transformation~\cite{lieb61} of
Hamiltonian~(\ref{eqn:hising}) leads to a Hamiltonian consisting
of two parts: a part that is quadratic in terms of Fermi operators
and a non-quadratic part:
\begin{eqnarray*}
    H/J  & = & -\sum_{i=0}^{N-1} \Big\{ \big( \adj{c_i} c_{i+1} +
    \adj{c_i} \adj{c_{i+1}} + \textrm{h.c.} \big) - 2 g \adj{c_i}
    c_i \Big\} +\\
    & & + \big( \adj{c_{N-1}} c_{0} + \adj{c_{N-1}} \adj{c_0} + \textrm{h.c.}
    \big) \big( e^{i \pi \hat{N}} + 1 \big) - g N
\end{eqnarray*}
The Fermi-operators~$c_j$ are chosen to be periodic, i.e. $c_{j+N}
\equiv c_j$. The symbol $\hat{N}$ denotes the fermion number
operator which is defined as $\hat{N} = \sum_{i=0}^{N-1} \adj{c_i}
c_i$. Even though the non-quadratic term can be neglected for many
calculations because it only makes changes of order~$\frac{1}{N}$
to eigenvalues and eigenstates~\cite{sachdev,pfeuty70}, it is
fundamental for the investigation of the quantum dynamics in the
vicinity of the critical point~$g=1$. This is because the ground
state and the lowest excited state approach according to~$\frac{4
\pi}{N}$ in the vicinity of the critical point as the number of
particles~$N$ increases (see figure~\ref{fig:myalgising:wn0}).
Thus, corrections of order~$\frac{1}{N}$ cannot be neglected even
if~$N$ is large.

Since the non-quadratic term~$e^{i \pi \hat{N}}$ is invariant
under linear transformations between
Fermi-operators~\cite{lieb61}, diagonalization can be performed
separately in two subspaces consisting of an even and an odd
number of fermions. In the odd-fermion-number subspace, $e^{i \pi
\hat{N}}$ simplifies to~$-1$ and the remaining quadratic
Hamiltonian can be diagonalized by linearly transforming the set
of Fermi-operators~$c_j$ into a set of
Fermi-operators~$\bar{\eta}_j$ in terms of which
Hamiltonian~(\ref{eqn:hising}) is diagonal:
\begin{equation} \label{eqn:hisingjwodd}
     H/J = g N + \sum_{k=0}^{N-1} \bar{\Lambda}_k \left( \adj{\bar{\eta}_k} \bar{\eta}_k - \frac{1}{2} \right)
\end{equation}
The energy~$\bar{\Lambda}_k$ of a single~$\bar{\eta}_k$-fermions
is calculated~as
\begin{equation} \label{eqn:lambdak}
    \bar{\Lambda}_k = 2 \left\{
    \begin{array}{lcl}
        g-1, & & k=0\\
        \sqrt{1 + g^2 - 2 g \cos \frac{2 \pi k}{N} }, & &
        k \neq 0\\
    \end{array}
    \right\}
    .
\end{equation}
In the even-fermion-number subspace $e^{i \pi \hat{N}}$ equals~$1$
and diagonalization yields
\begin{equation} \label{eqn:hisingjweven}
    H/J = g N + \sum_{k=0}^{N-1} \Lambda_k \left( \adj{\eta_k} \eta_k - \frac{1}{2} \right)
\end{equation}
with
\begin{equation} \label{eqn:lambdakq}
    \Lambda_k = 2 \sqrt{1 + g^2 + 2 g \cos \frac{2 \pi k}{N} }
\end{equation}
being the energy of one single $\eta_k$-fermion.

The ground state of Hamiltonian~(\ref{eqn:hising}) is the
vacuum-state $\ket{vac}$ in the even-fermion-number subspace. The
first excited state is the $1$-fermion state $\adj{\bar{\eta}_0}
\ket{vac}$ lying in the odd-particle subspace. The energy
difference between these two states amounts to
\begin{displaymath}
    \Delta E(g) = (g-1)+\sqrt{(g-1)^2+g \left( \frac{\pi}{N} \right)^2
    }
\end{displaymath}
for $N \gg 1$. This energy difference tends to zero for $g \leq 1$
in the thermodynamic limit, which is known as asymptotic
degeneracy~\cite{pfeuty70}.

\section{Calculating the Dynamics of the Ising Model} \label{sec:calculation_ising}

Adiabatic processes are conveniently treated by means of the
Adiabatic Approximation~\cite{messiah}. The Adiabatic
Approximation allows to calculate in first order the probability
that, during an adiabatic passage, states above the ground state
are excited. If it is assumed that the adiabatic process consists
in the variation of the parameter~$s$ from~$0$ to~$1$ during
time~$T$, the first order excitation probabilities read
\begin{equation} \label{eqn:adiabaticapprox2}
    p_{0 \to j}(T) \lesssim \frac{1}{T^2} \max_{0 \leq s \leq 1} \left|
    \frac{A_{j0}(s)}{\omega_{j0}(s)} \right|^2
\end{equation}
with
\begin{eqnarray*}
    A_{j0}(s) & = & - \frac{\bra{\psi_j(s)} \frac{d H}{d s} \ket{\psi_0(s)}}{\hbar \omega_{j0}(s)}\\
    \hbar \omega_{j0}(s) & = & E_j(s) - E_0(s).
\end{eqnarray*}
$\ket{\psi_j(s)}$ and $E_j(s)$ thereby denote the
$j^{th}$~Eigenstate of $H(s)$ and its corresponding Eigenvalue. A
definite statement about the duration of the adiabatic process is
finally obtained by demanding that the overall excitation
probability must be negligible, i.e. $p_E \equiv \sum_{j \neq 0}
p_{0 \to j}(T) \ll 1$.

In the adiabatic process investigated here, the field-strength~$g$
is varied linearly from~$g_0$ to $g_1$ during time~$T$. In terms
of the previously introduced formalism, $g$ is dependent on the
parameter~$s$ according to the formula
\begin{displaymath}
    g(s) = g_0 + s (g_1 - g_0)
\end{displaymath}
and $s$ is varied from $0$ to $1$ as time evolves from~$0$ to~$T$.

Since the adiabatic passage starts from the ground state (lying in
the even-fermion-number subspace) and the evenness and oddness of
states is conserved during time evolution because of the
$Z_2$-symmetry, calculations can be restricted to the
even-fermion-number subspace. In this subspace, the dynamics are
governed by Hamiltonian~(\ref{eqn:hisingjweven}).


The excitation probability can now be calculated by means of
formula~(\ref{eqn:adiabaticapprox2}). However, the determination
of the functions~$A_{j0}(s)$ requires the knowledge of the
derivative of $H(s)$ with respect to the parameter~$s$. This
derivative can be calculated easily if it is taken into account
that (for an odd number~$N$ of spins)
\begin{displaymath}
    \frac{d}{ds} \eta_k(s) = - \frac{\theta_k'(s)}{2} \left(
    \eta_{-k}(s) - \adj{\eta_{-k}}(s) \right),
\end{displaymath}
with
\begin{displaymath}
    \theta_k'(s) = \frac{d g}{d s} \, \frac{\sin \frac{2 \pi
    k}{N}}{1+g(s)^2+2 g(s) \cos \frac{2 \pi k}{N} }.
\end{displaymath}
The derivative $H'(s)$ couples the vacuum state $\ket{vac(s)}$
only to the vacuum state itself and to the two-fermion states
\begin{displaymath}
    \ket{n(s)} := \adj{\eta_n}(s) \adj{\eta_{-n}}(s)
    \ket{vac(s)},
    \quad
    n = 1, \ldots, \lfloor \frac{N-1}{2} \rfloor.
\end{displaymath}
The only states that are excited (in first order) during the
adiabatic passage are therefore the two-fermion states
$\ket{n(s)}$. The matrix element of $H'(s)$ that describes the
coupling between the states $\ket{n(s)}$ and $\ket{vac(s)}$ reads
\begin{displaymath}
    \bra{n(s)} H'(s) \ket{vac(s)} = - J \Lambda_n(s)
    \theta_n'(s).
\end{displaymath}
In addition, the energy difference between $\ket{n(s)}$ and the
ground state amounts to
\begin{displaymath}
    \hbar \omega_{n0}(s) = 2 J \Lambda_n(s).
\end{displaymath}
The probability that the two-fermion states $\ket{n(s)}$ are
excited can now be calculated using
formula~(\ref{eqn:adiabaticapprox2}):
\begin{displaymath}
    p_{0 \to n} \lesssim
    \frac{\hbar^2}{2^4 J^2} \frac{1}{T^2} \max_{0 \leq s \leq 1}
    \left| \frac{\theta_n'(s)}{\Lambda_n(s)} \right|^2
\end{displaymath}
The overall excitation probability~$p_E$, i.e. the probability
that any state above the ground state is excited during the
adiabatic process, is equal to the sum over all terms $p_{0 \to
n}$. This sum can be evaluated for $N \gg 1$ either by replacing
the sum by an integral yielding
result~(\ref{eqn:myalgising:peapproxa}) or by explicitly summing
over approximated addends yielding
results~(\ref{eqn:myalgising:peapproxb})
and~(\ref{eqn:myalgising:peapproxc}).

\section{Calculating the Dynamics of the Heisenberg Model} \label{sec:calculation_heisenberg}

As in the previous section, the probability that the system is, at
the end of the adiabatic process, in an excited state can be
calculation by means of formula~(\ref{eqn:adiabaticapprox2}). The
calculation of the quantities $A_{j0}(s)$ and $\hbar
\omega_{j0}(s)$, however, requires the knowledge of the
eigenstates and eigenvalues of
Hamiltonian~(\ref{eqn:hheisenberg}). The series expansions about
$g \to \infty$ of the ground state energy~$E_0(s)$ and the ground
state~$\ket{\psi_0(s)}$ read
\begin{displaymath}
    \frac{E_0(s)}{J N} = - g(s) - \Delta_x + \omicron{g}
\end{displaymath}
and
\begin{displaymath}
    \ket{\psi_0(s)} \propto \ket{\varphi_0^{(0)}} + \frac{1}{g(s)} \ket{\varphi_0^{(1)}} +
        \omicron{g}^2.
\end{displaymath}
The coefficients $\ket{\varphi_0^{(0)}}$ and
$\ket{\varphi_0^{(1)}}$ are thereby defined as
\begin{eqnarray*}
    \ket{\varphi_0^{(0)}} & = & \ketr\\
    \ket{\varphi_0^{(1)}} & = & \frac{1}{4} \left( \Delta_z -
        \Delta_y \right) \sum_{j=0}^{N-1} \ket{j,j+1}.
\end{eqnarray*}
The symmetries that the ground state possesses are the
$Z_2$-symmetry $1$ and the translational-symmetry $k=0$. The only
states that can be excited during a time evolution according to
$H(s)$ are states that possess the \emph{same} symmetries as the
ground state. The lowest states that meet this requirement are the
states with group-index $n=2$ and translational-symmetry $k=0$,
namely,
\begin{displaymath}
    \ket{\psi_{2 0 d}(s)}.
\end{displaymath}
The energies of these states are $E_{2 0 d}(s)$. In the case of an
odd number of particles $N$, the series expansions about $g \to
\infty$ of $E_{2 0 d}(s)$ and $\ket{\psi_{2 0 d}(s)}$ read
\begin{eqnarray*}
    \frac{E_{2 0 d}(s)}{J N} & = & - g(s) \left( 1- \frac{4}{N} \right) -
    \Delta_x \left( 1-\frac{8}{N} \right) +\\
    & & + \frac{4}{N} \left| \Delta_z + \Delta_y \right|
    \cos \frac{2 \pi d}{N+1}
    + \omicron{g}
\end{eqnarray*}
and
\begin{displaymath}
    \ket{\psi_{2 0 d}(s)} \propto \ket{\varphi_{2 0 d}^{(0)}} + \frac{1}{g(s)} \ket{\varphi_{2 0 d}^{(1)}} +
        \omicron{g}^2.
\end{displaymath}
The index $d$ may thereby assume the values
$1,\ldots,\frac{N-1}{2}$. The coefficients $\ket{\varphi_{2 0
d}^{(0)}}$ and the projection of $\ket{\varphi_{2 0 d}^{(1)}}$ on
$\ketr$ are given by
\begin{eqnarray*}
    \ket{\varphi_{2 0 d}^{(0)}} & = &
    \frac{2}{\sqrt{N+1}} \,
    \sum_{j=1}^{\frac{N-1}{2}}
    \sin \frac{2 \pi d j}{N+1} \ket{\Upsilon_{2 0 j}}\\
    \langle \rightarrow \ket{\varphi_{2 0 d}^{(1)}} & = & - \frac{1}{2}
    (\Delta_z - \Delta_y) \sqrt{\frac{N}{N+1}}
    \sin \frac{2 \pi d}{N+1}.
\end{eqnarray*}
The kets $\ket{\Upsilon_{2 0 j}}$ that appear in the upper formula
denote the $2$-particle states with translational symmetry $k=0$.
They are defined as
\begin{displaymath}
    \ket{\Upsilon_{2 0 j}} = \frac{1}{\sqrt{N}}
    \sum_{i=0}^{N-1} \ket{i,i+d}.
\end{displaymath}

In order to calculate the functions $A_{j0}(s)$, the matrix
elements of the derivative $H'(s)$ between the ground state
$\ket{\psi_0(s)}$ and the excited states $\ket{\psi_{2 0 d}(s)}$
must be determined. It is advantageous to abbreviate these matrix
elements by the notation $\HH_{d0}(s)$:
\begin{displaymath}
    \HH_{d0}(s) := \bra{\psi_{2 0 d}(s)} H'(s)
    \ket{\psi_0(s)}
\end{displaymath}
The derivative $H'(s)$ is equal to the expression
\begin{displaymath}
    H'(s) = - J g'(s) \sum_{i=0}^{N-1} \sx{i}.
\end{displaymath}
Using this expression and the series expansions of
$\ket{\psi_0(s)}$ and $\ket{\psi_{2 0 d}(s)}$, it is
straightforward to determine the matrix elements $\HH_{d0}(s)$:
The contribution to $\HH_{d0}(s)$ to $0^{th}$-order in
$\frac{1}{g}$ is zero. This is because the derivative $H'(s)$ does
not couple the $0^{th}$-order-coefficients $\ket{\varphi_0^{(0)}}$
and $\ket{\varphi_{2 0 d}^{(0)}}$. To $1^{st}$ order in
$\frac{1}{g}$, the terms $\bra{\varphi_{2 0 d}^{(1)}} H'(s)
\ket{\varphi_0^{(0)}}$ and $\bra{\varphi_{2 0 d}^{(0)}} H'(s)
\ket{\varphi_0^{(1)}}$ contribute. Higher-order contributions to
$\HH_{d0}$ will not be taken into account. In summary, the series
expansion to lowest order in $\frac{1}{g}$ of the matrix element
$\HH_{d0}$ reads
\begin{eqnarray*}
    \HH_{d0}(s) =
    \frac{1}{g(s)} & \Big\{ &
    \bra{\varphi_{2 0 d}^{(1)}} H'(s) \ket{\varphi_0^{(0)}} +\\
    & + &
    \bra{\varphi_{2 0 d}^{(0)}} H'(s) \ket{\varphi_0^{(1)}} \,
    \Big\} + \omicron{g}^2.
\end{eqnarray*}
A simplification of this expression yields
\begin{displaymath}
    \HH_{d0}(s) =
    -2 \left( \Delta_z - \Delta_y \right)
    \frac{g'(s)}{g(s)}
    \sqrt{\frac{N}{N+1}}
    \sin \frac{2 \pi d}{N+1} +
    \omicron{g}^2.
\end{displaymath}

The probability~$p_{0 \to d}$ that the states $\ket{\psi_{2 0
d}(s)}$ are excited can now easily be calculated using formula
(\ref{eqn:adiabaticapprox2}). The overall excitation
probability~$p_E$ is equal to the sum over all terms~$p_{0 \to
d}$. In the limit of a large number of particles ($N \gg 1$), this
sum can be approximated by an integral. This approximation leads
to a simplified expression for the overall excitation
probability~$p_E$, namely,
result~(\ref{eqn:myalgheisenberg:peapprox}).


\end{document}